\documentclass[a4paper]{jpconf}
\usepackage{graphicx}
\begin{document}
\title{Dimuons in pp and PbPb collisions with the CMS experiment }

\author{Jorge A. Robles}

\address{One Shields Ave. Physics Department, Davis CA, 95616 }

\ead{Robles@physics.ucdavis.edu}

\begin{abstract}
The unprecedented centre-of-mass energy available at the LHC offers unique opportunities for studying the properties of the strongly-interacting QCD matter created in PbPb collisions at extreme temperatures and very low parton momentum fractions. With its high precision, large acceptance for tracking and calorimetry, and a trigger scheme that allows analysis of each minimum bias PbPb event, CMS is fully equipped to measure dimuons in the high multiplicity environment of nucleus-nucleus collisions. Such probes are especially relevant for these studies since they are produced at early times and propagate through the medium, mapping its evolution. Inclusive and differential measurements of the Z boson yield will be shown, establishing that no modification is observed with respect to theoretical next-to-leading order pQCD calculations. The validity of the Glauber scaling for perturbative cross sections in nucleus-nucleus collisions at the LHC is confirmed. In addition, the status of the quarkonia measurements (primary and secondary J/$\psi$, and the 3 Upsilon states) will be reviewed, both in pp and in PbPb collisions. 
\end{abstract}

\section{Introduction}

The strongly interacting state of matter created in heavy-ion collisions can be cleanly probed via processes that decay in the dilepton channel. Given the fact that leptons do not interact via the strong force, these can traverse the medium unaffected. When studying dileptons experimentally, muons offer some advantages over electrons. The latter suffer energy losses from Bremsstrahlung radiation, while muons are much less susceptible to these effects. In the CMS experiment dimuon reconstruction offers a very good mass resolution for quarkonia and electro-weak processes. In pp collisions quarkonia  can be studied to test production models, while precision NNLO calculations can be tested with electroweak processes. The observables in pp act as a baseline to study the modification of probes seen in PbPb collisions. The study of quarkonia in heavy-ion collisions serves as a probe of the temperature of the Quark Gluon Plasma (QGP), which is expected to melt bound quarkonia states as the temperature rises. The study of Z bosons, since they are unaffected by the medium, are a very important tool to understand the cold nuclear effects. The energy density of the medium can also be measured via energy loss in Z+jet events, in which the Z balances the momentum (at tree level) of the jet on the opposite side.

\section{Muon reconstruction}

The CMS experiment is specially suited for muon reconstruction spanning a wide mass, with very good mass resolution. The reconstruction algorithm makes use of various detectors, combining the good muon identification and fast muon triggering from the muon chambers with the fine $p_{T}$ resolution from the inner tracker. The tracker, made up from pixels and silicon strips, has a coverage of $|\eta| <$ 2.5 with a $p_{T}$ resolution of 1-2$\%$ up to 100 GeV/c. The muon chambers, made up of 3 technologies, has a coverage of $|\eta| <$ 2.4. A full description of the CMS detector can be found in \cite{CMSdetector}.

\section{pp Results}
 CMS has accumulated enough statistics at $\sqrt{s}$ = 7 TeV  to measure the cross sections and kinematic distributions for J/$\psi$, $\Upsilon$ and Z to compare to different available models. The J/$\psi$, $\Upsilon$ and Z results in pp are explained in more detail in \cite{ppJpsi}, \cite{ppUpsilon} and \cite{ppZ} respectively.

\begin{figure}[h]
\begin{minipage}{20pc}
\includegraphics[width=20pc]{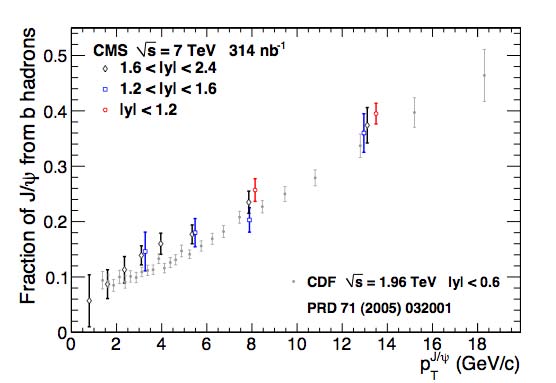}
\caption{\label{JpsiFrac} Fraction of J/$\psi$ from b-hadrons with data points from CMS at $\sqrt{s}$ = 7 TeV and CDF points at $\sqrt{s} $= 1.96 TeV.}
\end{minipage}
\hspace{2pc}
\begin{minipage}{16pc}
\includegraphics[width=16pc]{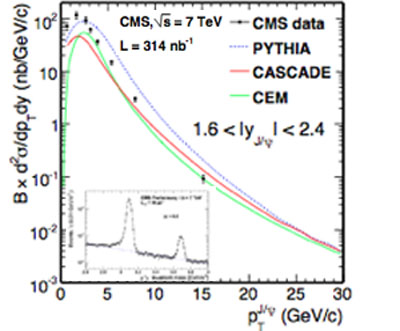}
\caption{\label{JpsiPtWithMass} J/$\psi$ $p_{T}$ distribution at forward rapidity, compared to models. Lower-left corner, J/$\psi$ and $\psi$' mass distribution }
\end{minipage}
\end{figure}

\subsection{$J/\psi$ in pp collisions}
The J/$\psi$ measurement requires proper identification of prompt J/$\psi$ and decay-in-flight contributions coming from b-hadrons. Measuring the fraction of prompt to non-prompt relies in the discrimination of J/$\psi$ mesons produced away from the collision vertex.  Fig.~\ref{JpsiFrac} shows the fraction of J/$\psi$ from $b$-hadron in three rapidity ranges as a function of transverse momentum.  The measured fraction of J/$\psi$ from $b$-hadron obtained by CMS lies on top of  the data points obtained from the CDF experiment, regardless of the different center-of-mass energies. The lower-left plot on Fig.~\ref{JpsiPtWithMass} shows the invariant mass of the J/$\psi$ and the $\psi$' in the dimuon channel. The mass resolution observed is $\sigma$ = 20 MeV/$c^{2}$ in the rapidity range $|y| <$0.5. The $p_{T}$ spectra for  prompt and non-prompt J/$\psi$ is also measured in three rapidity bins and compared to available models (Color Evaporation Model, Pythia and CASCADE). In the main frame of Fig~\ref{JpsiPtWithMass} the $p_{T}$ spectra of J/$\psi$ at high rapidity is shown. A reasonable agreement is found between data and theory in all bins except at low $p_{T}$  at forward rapidity for the prompt J/$\psi$. The non-prompt J/$\psi$ is divided also in the same rapidity bins and compared to theory (CASCADE, Pythia and a B$\rightarrow$ J/$\psi$ FONLL calculation) in all bins a good agreement is found.


\subsection {$\Upsilon$ in pp collisions}

\begin{figure}[h]
\begin{minipage}{22pc}
\includegraphics[width=22pc]{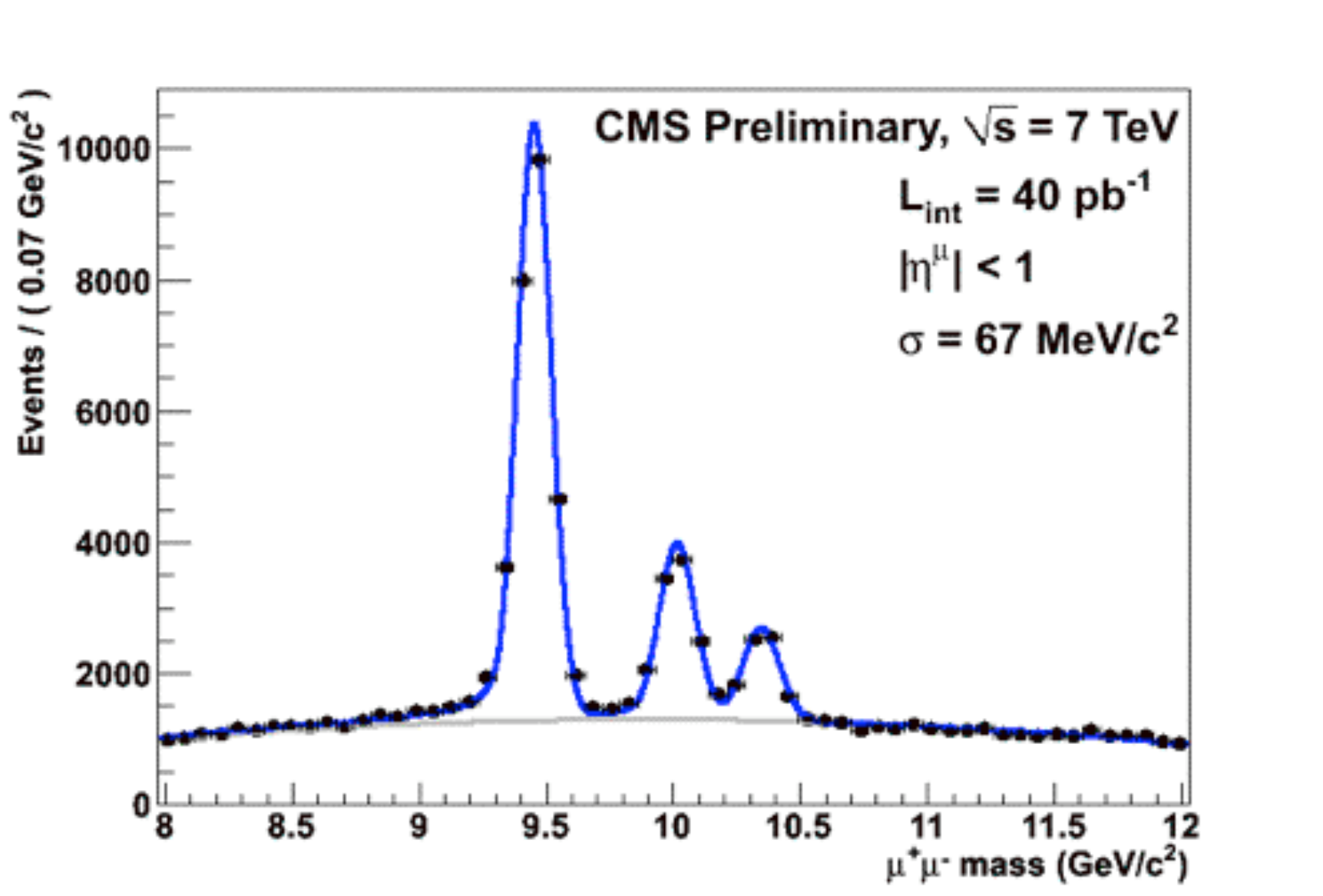}
\caption{\label{UpsilonMassBarrel}$\Upsilon$ states reconstructed in the CMS barrel $|\eta^{\mu}| <$1.0} 
\end{minipage}\hspace{2pc}%
\begin{minipage}{18pc}
\includegraphics[width=15pc]{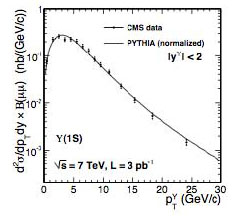}
\caption{\label{Upsilon1sPt}$\Upsilon$ 1S $p_{T}$ spectra, compared to PYTHIA in the $|y^{\Upsilon}|<2.0$ range} 
\end{minipage} 
\end{figure}

The superb mass resolution achieved by CMS  allows for a clear separation of the $\Upsilon$, $\Upsilon$' and $\Upsilon$''  states. The reconstructed mass resolution obtained over the full CMS acceptance, $|\eta^{\mu}| <$2.4, is $\sigma$ = 100 MeV/$c^{2}$. This resolution can be further improved to $\sigma$ = 67 MeV/$c^{2}$ by restricting to muons reconstructed in the barrel region, $|\eta^{\mu}| <$1.0 shown in Fig~\ref{UpsilonMassBarrel}. The differential cross-section down to zero $p_{T}$ for the 3 $\Upsilon$ states at $\sqrt{s}$ = 7 TeV was obtained. In Fig.~\ref{Upsilon1sPt} the $\Upsilon$(1S) $p_{T}$ spectra is shown and compared to Pythia in the $|y^{\Upsilon} |< $ 2 rapidity range assuming unpolarized $\Upsilon$(ns) production. For the 3 $\Upsilon $ states a good agreement is found within statistical uncertainties. The $\Upsilon$(1s) and $\Upsilon$(2s) include feed-down from $\chi_{b}$ and $\Upsilon$(3s). 

\subsection{Z in pp collisions}

\begin{figure}[h]
\begin{minipage}{15pc}
\includegraphics[width=15pc]{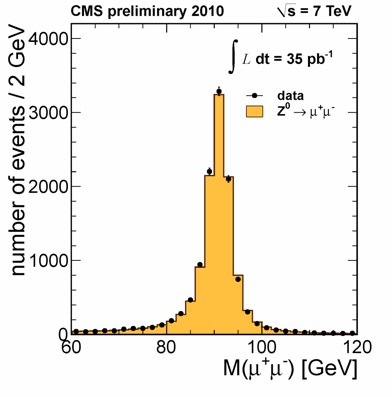}
\caption{\label{ZMassInpp} Z Invariant mass at $\sqrt{s}$ = 7 TeV data points  with MC histogram in yellow.}
\end{minipage}\hspace{1pc}%
\begin{minipage}{22pc}
\includegraphics[width=22pc]{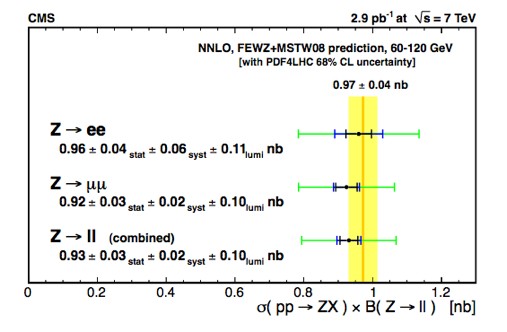}
\caption{\label{ZMeasuremntsInpp} Ratio of Z cross-section measurement in the di-lepton channels with NNLO calculations. } 
\end{minipage} 
\end{figure}

Z boson production has been studied to a high degree of precision in proton-proton experiments. Simulation is able to accurately describe the data obtained by CMS in the Z$\rightarrow \mu^{+} \mu{-}$ channel at  $\sqrt{s}$ = 7 TeV. Next-to-Next Leading Order(NNLO) calculations are in superb agreement with Z measurements in CMS. The Z cross-section measured in the di-electron and dimuon channel are compared to a NNLO calculation shown in Fig~\ref{ZMeasuremntsInpp}, the experimental uncertainties are  $<$ 4 $\%$. The comparison uncertainties are dominated by the luminosity error bars  ($\approx$11 $\%$), which are expected to diminish in the future.

\section{Heavy Ions in CMS}

The Heavy Ion program at the LHC represents an increase in the center of mass energies of a factor of 14 compared to RHIC. This allows us to study the QGP with probes that were previously inaccessible. The production rate at the LHC allows for precision bottomonia studies, jets with energy above 100 GeV are be available to probe the medium, and is the first time electroweak processes are available in heavy ion collisions. Overall, the high-$p_{T}$ regime plays an important role in measurements that aim to quantify medium effects. CMS is ideal to carry out measurements in the dimuon channel, taking a advantage of the high resolution of muon and tracker detectors that have a performance that is just as good as in pp collisions. The results shown in PbPb collisions were obtained with the CMS experiment during the first HI run at the LHC at $\sqrt{s}$ = 2.76 TeV, with data collected over November-December 2010.

\begin{figure}[h]
\begin{minipage}{18pc}
\includegraphics[width=18pc]{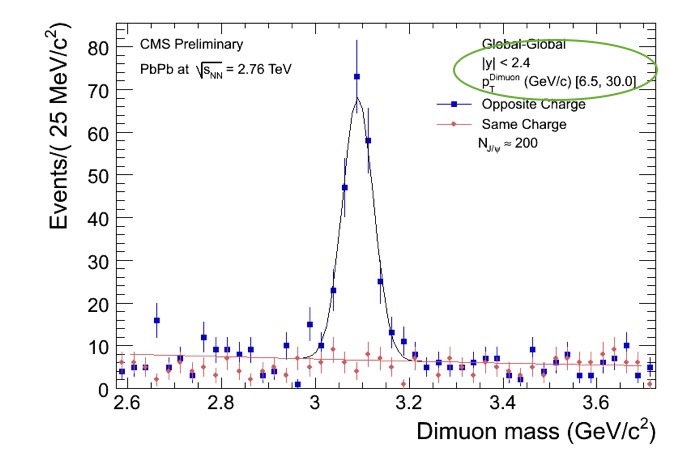}
\caption{\label{JpsiInPbPb} J/$\psi$ Invariant mass t $\sqrt{s}$ = 2.76 TeV in dimuon channel. Opposite sign pairs in blue and same-sign pairs in red. }
\end{minipage}\hspace{2pc}%
\begin{minipage}{18pc}
\includegraphics[width=18pc]{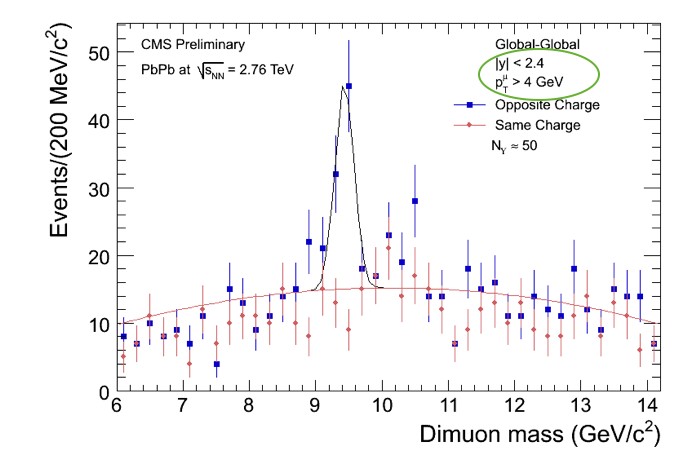}
\caption{\label{UpsilonInPbPb} $\Upsilon$ Invariant mass t $\sqrt{s}$ = 2.76 TeV in dimuon channel. Opposite sign pairs in blue and same-sign pairs in red.} 
\end{minipage} 
\end{figure}

Preliminary results obtained by a quick analysis are shown in Fig.~\ref{JpsiInPbPb} and Fig.~\ref{UpsilonInPbPb}, in which the invariant mass distribution around the J/$\psi$ and $\Upsilon$, respectively, are obtained in the dimuon channel. In the case of the J/$\psi$, the reconstructed muons are restricted to have a $p_{T}^{\mu} >$ 6.5 GeV/c, and the dimuon is restricted to be in the rapidity range $|y^{J/\psi}| < $ 2.4. A total of $\approx 200 $ J/$\psi$ counts is found in a peak of opposite sign pairs over the same-sign flat background. Fig~\ref{UpsilonInPbPb} shows the reconstructed invariant mass peak that arises from muons with a $p_{T}^{\mu} >$ 4 GeV/c cut, and in the dimuon rapidity region $|y^{\Upsilon}| < $ 2.4. These plots were obtained from a subset of heavy ion events available at the time.

\subsection{Quarkonia in PbPb collisions} 
\subsection{Z Results in PbPb collsions}

The Z $\rightarrow \mu^{+} \mu{^-}$ process in PbPb collisions offers a benchmark probe of the initial quark pdf in the Pb nucleus that can be cleanly extracted. The Z is produced from hard collisions via $q\bar{q} \rightarrow Z$ process, once produced it decays within the medium on a time scale of $\approx 10^{-24}$ seconds. The dimuon decay channel is of special interest because the Z, as well as the muons, traverse the medium unaffected by the strongly interacting Quark Gluon Plasma. Given that the entire process is virtually untouched by the medium it offers a probe to measure initial state effects. 

Predictions\cite{Ramona, Kart, Neufeld, Paukkunen, Zhang} indicate that the effects that can affect the Z boson production in heavy ion collisions are rather small. From these effects,  shadowing is expected to have the biggest contribution $\approx$10 - 20$\%$ \cite{Paukkunen}. The isospin effect arises from the difference of quark content of protons and neutrons that collide in the PbPb case as opposed to pp, and is expected to contribute less tha 3$\%$ \cite{Paukkunen}. Finally, scattering and energy loss of the initial partons should have an effect of the order of 2 $\%$\cite{Neufeld}.

\begin{figure}[h]
\begin{minipage}{40pc}
\hspace{6pc}
\includegraphics[width=26pc]{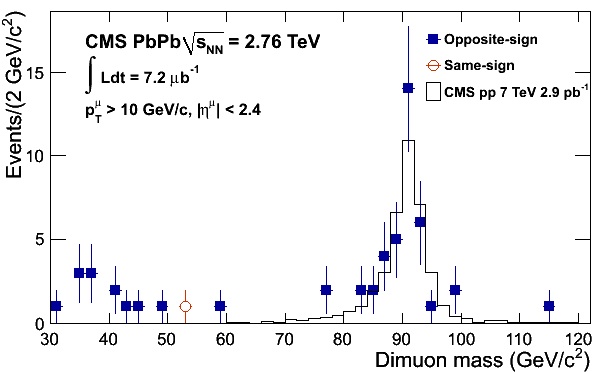}
\caption{\label{ZMassPbPb} Z invariant mass in PbPb collisions at $\sqrt{s}$ = 2.76 TeV in dimuon channel}
\end{minipage}
\end{figure}

The first measurement of Z production in PbPb collisions in CMS is further detailed in \cite{PbPbZ}. The reconstruction of the Z boson in the dimuon channel is done my requiring two opposite-charge muons with $p_{T}^{\mu} > $ 10 GeV/c and $|\eta^{\mu}| < 2.4$. A loose set of quality cuts applied in each of the muons is enough to extract a clear peak in the 30 - 120 GeV/$c^{2}$ mass regime. Fig.~\ref{ZMassPbPb} shows invariant mass from opposite-charge muon pairs in blue squares and only one same-sign pair in the above mentioned mass regime. The data points are overlaid with the empty histogram from pp collision data normalized to an integral of 39 counts. It is evident that Z $\rightarrow \mu^{+} \mu{^-}$ process can be extracted almost background free in a wide mass range around the Z mass. The structure found in the 30 - 50 GeV/$c^{2}$, is due to the dimuon continuum from other physics processes, mainly $b\bar{b}$ production.  A visible agreement between the PbPb data points and the pp histogram data points to a comparable detector performance in pp and heavy ion collisions.

\begin{figure}[h]
\begin{minipage}{12pc}
\includegraphics[width=12pc]{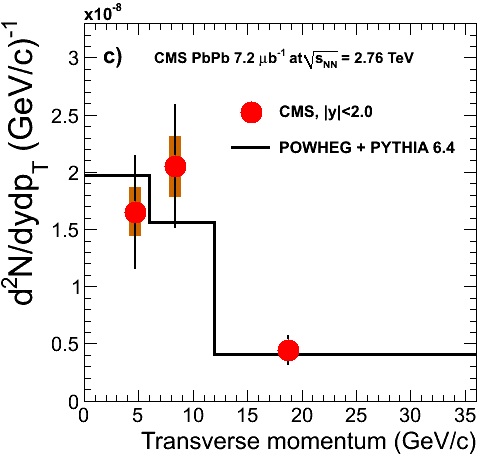}
\caption{\label{ZptPbPb} Z differential yield as a function of $p_{T}$}
\end{minipage}\hspace{2pc}%
\begin{minipage}{12pc}
\includegraphics[width=12pc]{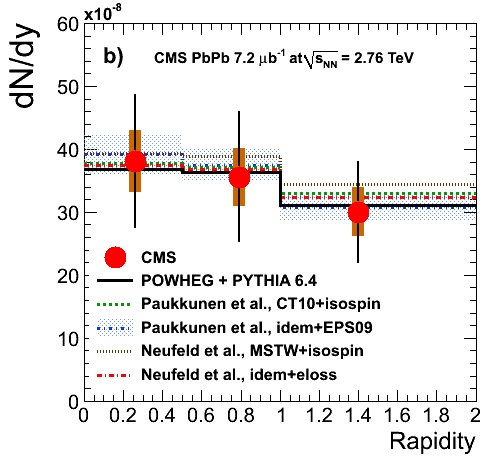}
\caption{\label{ZrapPbPb} Z differential yield as a function of rapidity } 
\end{minipage} \hspace{2pc}%
\begin{minipage}{12pc}
\includegraphics[width=12pc]{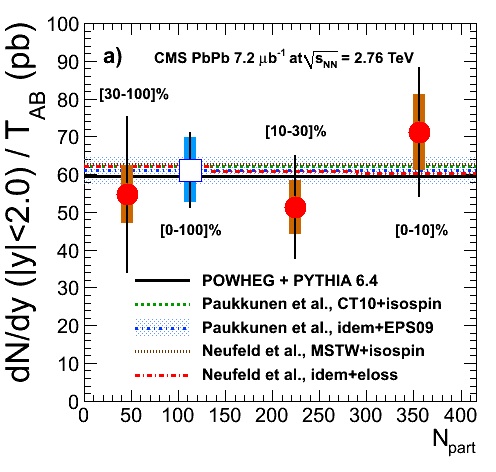}
\caption{\label{ZRaaPbPb} Z differential yield divided by the nuclear overlap function as a function of $N_{part}$} 
\end{minipage} 
\end{figure}

Given the number of candidate events obtained from the run, the data were subdivided into transverse momentum($p_{T}^{Z}$), rapidity($y^{Z}$) and event-centrality bins. In absence of in-medium modification Z boson production is expected to scale by the number of binary collisions with respect to the pp measurement. Figure~\ref{ZptPbPb} shows the Z differential yield as a function of transverse momentum. The statistical error bars are shown in black and the systematic uncertainties are shown in orange boxes. The red data points in the rapidity region $|y^{Z}| <$ 2.0 are compared to a next-to-leading order POWHEG calculation scaled by the nuclear geometry ($A^{2}/\sigma_{PbPb}$). Within statistical uncertainties an agreement between data and theory is found.

Figure~\ref{ZrapPbPb} shows the Z differential yield as a function of rapidity in three bins. The red data points, with statistical error bars and systematic uncertainties in boxes, are compared to various theoretical predictions scaled by the nuclear geometry. The black line is a POWHEG calculation interfaced with Pythia as a baseline. To further discern between initial state effects a few more models are included. With the CT10 parameterization, the green dashed line shows isospin effects, the blue dot-and-dash line shows isospin and shadowing effects,. With the MSTW parameterization, the dotted brown line shows isospin effects ,while the red dot-and-dash line shows isospin and energy loss effects. Within the measurement uncertainties none of these subtler effects can be excluded.

The differential yield divided by nuclear overlap function, $T_{AB}$, as a function of $N_{part}$ is shown in Fig~\ref{ZRaaPbPb} in three centrality bins and a minbias point. The $N_{part}$ value for each of the bins is obtained by the average number of participants in the given bin. The CMS data points, in the rapidity range $|y^{Z}| <$ 2.0, are compared to the same theoretical models described in the previous paragraph. The quantity plotted can be related to the nuclear modification factor, $R_{AA}$, by the following equation:
\begin{equation}
\frac{dN_{AA}}{dy} / T_{AB} = R_{AA} \times \frac{d\sigma_{pp}}{dy}
\end{equation}

The main feature is a flat distribution as a function of centrality. This is expected for a probe that is unmodified by the presence of the dense QGP produced in central collisions. The data is also observed to match the POWHEG pp simulation scaled by the nuclear geometry.

\section{Conclusions}

An overall agreement between theory and data for the J/$\psi$, $\Upsilon$ production is observed in pp collisions. A high level of agreement found for Z production measured in the dilepton channel and NNLO calculations. The J/$\psi$ and $\Upsilon$ were observed by CMS in PbPb collisions. The Z boson differential cross-sections, as a function of $y^{Z}$ and $p_{T}^{Z}$, were measured and observed to scale with the nuclear geometry with respect to theoretical NLO pQCD pp calculations. The Z yield as a function of centrality was found be flat, consistent an unmodified probe in central collisions.


\section*{References}
\begin{thebibliography}{11}
\bibitem{CMSdetector} R. Adolphi et al. (CMS), JINST 3, S08004 (2008)
\bibitem{ppJpsi}CMS collaboration, Eur. Phys. J. C (2011) 71: 1575, arXiv:1011.4193 [hep-ex].
\bibitem{ppUpsilon}CMS collaboration, arXiv:1012.5545v1 [hep-ex]
\bibitem{ppZ}CMS collaboration, JHEP01 (2011) 080, arXiv1012.5545 [hep-ex].
\bibitem{Ramona} R. Vogt, Phys. Rev. C64, 044901 (2001), arXiv:hep-ph/0011242.
\bibitem{Kart}V. Kartvelishvili, R. Kvatadze, and R. Shanidze, Phys.Lett. B356, 589 (1995), arXiv:hep-ph/9505418.
\bibitem{Neufeld}R. B. Neufeld, I. Vitev, and B. W. Zhang, (2010),arXiv:1006.2389 [hep-ph].
\bibitem{Paukkunen}H. Paukkunen and C. A. Salgado, (2010),arXiv:1010.5392 [hep-ph].
\bibitem{Zhang}X.-F. Zhang and G. I. Fai, Phys. Lett. B545, 91 (2002),arXiv:hep-ph/0205155.
\bibitem{PbPbZ}CMS collaboration, PRL106, 212301 (2011), arXiv: 1102:5435 [nucl-ex].
\end{thebibliography}
\end{document}